\documentclass[11pt]{article}
\usepackage{amssymb}
\usepackage[T1]{fontenc}
\usepackage[latin2]{inputenc}
\usepackage{babel}

\makeatletter
\linethickness{0.4pt}

\begin{document}
\newtheorem{lema}{Lemma}
\newtheorem{theorem}{Theorem}
\newtheorem{corollary}{Corollary}
\newtheorem{proposition}{Proposition}
\newtheorem{example}{Example}
\newtheorem{proof}{Proof}

\ \hfill IFT UwB /13/2000

\ \hfill November 18, 2000

\vspace*{1cm}

\begin{center}

{\huge\bf The Darboux Transform and some Integrable cases of the $q$-Riccati Equation\footnote { Supported in part by KBN grant 2 PO3 A 012 19}}\\

\vspace*{1cm}

{\bf Anatol Odzijewicz{*} \& Alina Ryżko{*}{*}}\\

\vspace*{1cm}

Institute of Theoretical Physics\\

University in Białystok\\

ul. Lipowa 41, 15-424 Białystok, Poland\\

E-mail: {*}aodzijew@labfiz.uwb.edu.pl, {*}{*}alaryzko@alpha.uwb.edu.pl \\

\vspace*{1cm}

\end{center}

\begin{abstract}
\noindent
{ Using the q-version of the Darboux transform we obtain
the  general solution of $q$--difference Riccati equation from 
a special one by the action of one--parameter group. This allows us to construct the solutions
for the large class of $q$--difference Riccati equations as well as 
$q$--diference Schr{\"o}dinger equations.}
\end{abstract}
\vspace{1.5cm}
\thispagestyle{empty}

\section*{Introduction}

In this paper we investigate the Darboux factorization method for $q$--difference 
version of Riccati and Schr{\"o}dinger equations. It appears that this method which 
is by all means effective for differential Riccati and Schr{\"o}dinger equation (\cite{7,8,2}) leads 
to non--trival and interesting results
in the $q$--deformed case too. Some of the new formulae have their undeformed version.
They tend in the limit of $q\rightarrow 1$ to the ones, which are well known in differential case.
\vspace*{0.2cm}

The Darboux factorization
\begin{equation}
\label{1}
-\frac{d^{2}}{dx^{2}}+V(x)=\left( \frac{d}{dx}+u(x)\right) \left( -\frac{d}{dx}+u(x)\right) 
\end{equation}
gives  the well known correspondence between one-dimensional Schr{\"o}dinger equation
\begin{equation}
\label{1a}
\left( -\frac{d^{2}}{dx^{2}}\psi (x)+V(x)\right) \psi (x)=0
\end{equation}
and the Riccati equation
\begin{equation}
\label{2}
\frac{d}{dx}u(x)=-u^{2}(x)+V(x),
\end{equation}
where 
\begin{equation}
\label{3}
u(x)=\frac{d}{dx}ln\psi (x).
\end{equation}

This correspondence is a starting point for a search of the exact solutions of
both equations above (see \cite{2}).

We will investigate, in this paper, an analogue of the Darboux method for the
pair of $q$-difference equation
\begin{equation}
\label{5}
\partial _{q}\left( \begin{array}{c}
\psi (x)\\
\varphi (x)
\end{array}\right) =\left( \begin{array}{cc}
R(x) & S(x)\\
V(x) & T(x)
\end{array}\right) \left( \begin{array}{c}
\psi (x)\\
\varphi (x)
\end{array}\right) ,
\end{equation}
and
\begin{equation}
\label{4}
V(x)=\partial _{q}u(x)-T(x)u(x)+R(x)u(qx)+S(x)u(x)u(qx).
\end{equation}
whose solutions are related by
\begin{equation}
\label{6}
u(x)=\frac{\varphi (x)}{\psi (x)}.
\end{equation}
It is clear that (\ref{5}) and (\ref{4}) generalize (\ref{1a}) and (\ref{2})
respectively. The Schr{\"o}dinger and Riccati equations are obtained in the
limit \( \:\: q\rightarrow 1 \) under additional assumption that \( \: R(x)=0=T(x) \). 

Let us recall here, that the $q$-derivative and $q$-integral are defined by

\begin{equation}
\label{7}
\partial _{q}\psi (x)=\frac{\psi (x)-\psi (qx)}{(1-q)x},
\end{equation}

\begin{equation}
\label{7a}
\int ^{x}_{0}\psi (x)d_{q}t=\sum _{n=0}^{\infty }(1-q)q^{n}x\psi (q^{n}x)
\end{equation}
respectively, where \( \: 0\leq q\leq 1\:  \). The standard derivative and
integral are obtained for \( \: q=1 \). However, the reason for the investigation of 
the $q$-difference equations
(\ref{5}) and (\ref{4}) is not only that they generalize in a natural way 
the Schr{\"o}dinger and Riccati equations.

If one additionally assumes \( \: 1-(1-q)xT(x)=0 \)
in the real case and one takes \( \: q^{n} \) instead of  the real argument
\( \: x \), the equation (\ref{5}) appear to reduce to the three terms requrence
equation
\begin{equation}
\label{8}
\psi _{n+2}=\left[1-(1-q)q^{n+1}R(q^{n+1})\right]\psi _{n+1}+(1-q)^{2}q^{2n+1}S(q^{n+1})V(q^{n})\psi _{n},
\end{equation}
for the function \( \: \psi _{n}:=\psi (q^{n}) \) of the natural argument $n\in{I\!\!N}\cup\{0\}$.
Hence, the $q$-difference equation (\ref{5}) can be applied to those physical
problems which are related to the theory of orthogonal polynomials (see \cite{4}) .

The paper is organized in the following way.
In Section 1 we will introduce
the $q$-difference Darboux transform and we will integrate the equation (\ref{5})
for the case of \(  V(x)=0 \). The action of $q$--difference Darboux transform
will be presented in Section 2. There we find an one--parameter auto-B{\"a}cklund transform
for the $q$-difference Riccati equation and we show that it generates the general
solution of (\ref{4}) from a special one. Section 3 is devoted to the presentation
of some extendend classes of solutions of the $q$--difference Schr{\"o}dinger and Riccati  equation. All
results presented in Section 1 and  2 have well known differential counterparts
and this aspect is also exhibited in the paper.

\section{The $q$-difference Darboux transform}

In order to solve the $q$--difference equation (\ref{5}) by the iterative method
we will rewrite it in the following form
\begin{equation}
\label{9}
\left( \begin{array}{c}
\psi (qx)\\
\varphi (qx)
\end{array}\right) =\Lambda \left( x\right) \left( \begin{array}{c}
\psi (x)\\
\varphi (x)
\end{array}\right) ,
\end{equation}
where
\begin{equation}
\label{10}
\Lambda (x)={1\!\!\mbox{\rm I}} -(1-q)x\left( \begin{array}{cc}
R(x) & S(x)\\
V(x) & T(x)
\end{array}\right) .
\end{equation}
Let us assume here that $R(x)$, $S(x)$, $V(x)$ and $T(x)$
are continuous functions of a real argument. Hence, the matrix sequence
\begin{equation}
\label{11}
\Lambda (q^{n-1}x).....\Lambda (qx)\Lambda (x)=:\! \Lambda (x;q)_{n}
\end{equation}
is pointwise convergent
\begin{equation}
\label{12}
\Lambda (x;q)_{n}\longrightarrow_{n\rightarrow \infty }\Lambda (x;q)_{\infty }
\end{equation}
to a matrix function \( \! \Lambda (x;q)_{\infty } \). The inverse matrix function $\Lambda (x;q)_{\infty }^{-1} $
is exactly the resolvent of equation (\ref{5}), i. e.
\begin{equation}
\label{14}
\left( \begin{array}{c}
\psi (x)\\
\varphi (x)
\end{array}\right) =\Lambda (x;q)_{\infty }^{-1}\left( \begin{array}{c}
\psi (0)\\
\varphi (0)
\end{array}\right) .
\end{equation}
The problem of solving  (\ref{5}) thus  is equivalent to the calculation of the
infinite matrix product
\begin{equation}
\label{15}
\Lambda (x;q)_{\infty }^{}:=\prod _{k=0}^{\infty }\Lambda (q^{n}x).
\end{equation}
The above suggests the following transform:

\begin{equation}
\label{16a}
\Lambda (x)\quad \longrightarrow \quad D(qx)^{-1}\Lambda (x)D(x)=\Lambda ^{'}(x),
\end{equation}

\begin{equation}
\label{16}
\left( \begin{array}{c}
\psi (x)\\
\varphi (x)
\end{array}\right) \quad \longrightarrow \quad D(x)^{-1}\left( \begin{array}{c}
\psi (x)\\
\varphi (x)
\end{array}\right) =\left( \begin{array}{c}
\psi ^{'}(x)\\
\varphi ^{'}(x)
\end{array}\right) ,
\end{equation}
where \( \! D(x) \) is $GL(2,{I\!\!R})$ - valued function of the real argument. It is obvious that the transform (\ref{16a}-\ref{16}) preserves
the form of the equation (\ref{9}) and the transformed resolvent \( \Lambda ^{'}(x;q)_{\infty }^{-1} \)
is related to the initial one by
\begin{equation}
\label{17}
\Lambda ^{'}(x;q)_{\infty }^{-1}=D(x)^{-1}\Lambda (x;q)_{\infty }^{-1}D(0).
\end{equation}
Thus the virtue of the above transform is to find such matrix valued function
\( D(x) \) for equation (\ref{5}), which reduces unknown resolvent \( \Lambda (x;q)_{\infty }^{-1} \)
to some known one \( \Lambda ^{'}(x;q)_{\infty }^{-1} \).

We will find later the explicit form of the resolwent \( \Lambda (x;q)^{}_{\infty } \)
in the case when \mbox{\( V(x)=0 \)}. So, in order to integrate (\ref{5})
it is enough to transform (\ref{15}) to the uppertriangular matrix function
\( \Lambda (x) \) by the use of (\ref{16a}-\ref{16}). Any matrix can be
decomposed generically into the product of the uppertriangular and lowertriangular
matrices. Thus, without loss of generality we can assume that 
\begin{equation}
\label{19}
D(x)=\left( \begin{array}{cc}
1 & 0\\
-c(x) & 1
\end{array}\right) .
\end{equation}
After substituting (\ref{19}) into (\ref{16a}) we find that \( \: \Lambda ^{'}(x) \)
will be an uppertriangular matrix if and only if the function \( \: c(x) \)
satisfies the $q$--difference Riccati equation (\ref{4}).

The $q$--difference Schr{\"o}dinger operator factorizes into the form:
\begin{equation}
\label{19a}
-\partial _{q}^{2}+V(x)=\left( \partial _{q}+u(qx)\right) \left( -\partial _{q}+u(x)\right) ,
\end{equation}
iff the function $u(x)$ satisfies the equation (\ref{4}) with
$R(x)=T(x)=0$  and $S(x)=1$.
Hence, it is natural to call the matrix transform (\ref{16a}-\ref{16}) {\bf the
$q$--difference Darboux transform}.

We will use the identity
\begin{equation}
\label{19b}
\prod _{n=0}^{\infty }\left( 1-(1-q)q^{n}xf(q^{n}x)\right) =\exp \left( \frac{1}{1-q}\int ^{x}_{0}\frac{\ln \left( 1-(1-q)tf(t)\right) }{t}d_{q}t\right) ,
\end{equation}
which is an easy consequence of the definition of  $q$--integral.

\vspace*{0.2cm}

\begin{proposition}
If \( R(x) \), \( S(x) \) and \( T(x) \) are continuous functions and \mbox{\( V(x)=0 \)}
then
\begin{equation}
\label{21}
\Lambda (x;q)_{\infty }=\left[ \begin{array}{cc}
\exp \frac{1}{1-q}\int _{0}^{x}\frac{\ln (1-(1-q)tR(t))}{t}d_{q}t & B(x)\\
0 & \exp \frac{1}{1-q}\int _{0}^{x}\frac{\ln (1-(1-q)tT(t))}{t}d_                                                                                                                                   {q}t
\end{array}\right] ,
\end{equation}
where  
\[
B(x)=\left[-\exp \left( \frac{1}{1-q}\int _{0}^{x}\frac{\ln (1-(1-q)tT(t))}{t}d_{q}t\right)\right] \times 
\]
\begin{equation}
\label{21a}
\times \left[ \int _{0}^{x}\frac{S(t)}{1-(1-q)tR(t)}\exp \left( \frac{1}{1-q}\int _{0}^{t}\frac{1}{s}\ln \frac{1-(1-q)sR(s)}{1-(1-q)sT(s)}d_{q}s\right) d_{q}t\! .\right] \: .
\end{equation}
\end{proposition}

\vspace*{0.2cm}\hspace*{-0.7cm}
{\it Proof:}\\
Since \( \: V(x)=0 \) we can assume that \( \Lambda (x;q)_{\infty } \) is the
uppertriangular matrix of the form 
\begin{equation}
\label{22}
\Lambda (x;q)_{\infty }=\left[ \begin{array}{cc}
\prod ^{\infty }_{n=0}\left( 1-(1-q)q^{n}xR(q^{n}x)\right)  & B(x)\\
0 & \prod _{n=0}^{\infty }\left( 1-(1-q)q^{n}xT(q^{n}x)\right) 
\end{array}\right] .
\end{equation}
where from  the equation
\begin{equation}
\label{22a}
\Lambda (x;q)_{\infty }=\Lambda (qx;q)_{\infty } \Lambda (x) ,
\end{equation}
we find that the function \( B(x) \) does satisfy 
\begin{equation}
\label{24}
B(x)=\left[ 1-(1-q)xT(x)\right] B(qx)-\prod ^{\infty }_{n=0}\left( 1-(1-q)q^{n+1}xR(q^{n+1}x)\right) \left( 1-q\right) xS(x).
\end{equation}
The equation ( \ref{24} ) is solved by iterative method. One gets finally:
\[
B(x)=-\prod ^{\infty }_{n=0}\left( 1-(1-q)q^{n}xT(q^{n}x)\right) \times 
\]
\begin{equation}
\label{25}
\times \sum ^{\infty }_{n=0}\frac{\left( 1-q\right) q^{n}xS(q^{n}x)}{1-\left( 1-q\right) q^{n}xR(q^{n}x)}\prod _{k=n}^{\infty }\frac{\left( 1-(1-q)q^{k}xR(q^{k}x)\right) }{\left( 1-(1-q)q^{k}xT(q^{k}x)\right) }\: .
\end{equation}
Substituting (\ref{25}) into (\ref{22}) and using identity (\ref{19b})
we obtain the formulae (\ref{21}) and (\ref{21a}).

\ \hfill QED

\vspace*{0.2cm}

\section {The solution of $q$-difference Schr{\"o}dinger equation and auto-B{\"a}clund
transform for $q$-difference Riccati equation}

We have obtained in Section 1, the resolvent function \( \Lambda (x;q)_{\infty }^{-1}\: \)
for the case of \( \: V(x)=0 \), (see {\bf Proposition 1}). Let us assume \(  \Lambda ^{'}(x) \)
in (\ref{16a}) to be uppertriangular matrix function. Applying the $q$--difference
Darboux transform to \( \Lambda ^{}(x) \), with \( \: D(x) \) given by (\ref{19}),
we find the general solution of the $q$--differential equation (\ref{5}).

\vspace*{0.2cm}

\begin{proposition}
Let us assume that functions \(  R(x) \), \(  S(x) \), \(  T(x) \) and \(  V(x) \)
from the equations (\ref{5}) are continuous. Let \( \: \left( \begin{array}{cc}
\psi _{0}(x) & \varphi _{0}(x)
\end{array}\right) ^{\top } \) be a particular solution of (\ref{5}) one and put
\begin{equation}
\label{26a}
u_{0}(x)=\frac{\varphi _{0}(x)}{\psi _{0}(x)}\: .
\end{equation}
Then the potential \( \: V(x) \) is given by 
\begin{equation}
\label{26}
V(x)=\partial _{q}u_{o}(x)-T(x)u_{0}(x)+R(x)u_{0}(qx)+S(x)u_{0}(x)u_{0}(qx),
\end{equation}
and the general solutions  of (\ref{5}) is given by:
\[
\psi (x)=\exp \left( -\frac{1}{1-q}\int _{0}^{x}\frac{\ln \left\{ 1-(1-q)t\left[ R(t)+u_{0}(t)S(t)\right] \right\} }{t}d_{q}t\right) \times 
\]
\[
\times \left[ D+F\int _{0}^{x}\frac{S(t)}{1-\left( 1-q\right) t\left[ R(t)+u_{0}(t)S(t)\right] }\right. \times 
\]
\begin{equation}
\label{27}
\times \left. \exp \left( \frac{1}{1-q}\int ^{t}_{0}\frac{1}{s}\ln \frac{1-(1-q)s\left[ R(s)+u_{o}(s)S(s)\right] }{1-(1-q)s\left[ T(s)-u_{0}(qs)S(s)\right] }d_{q}s\right) d_{q}t\right] ,
\end{equation}
\[
\varphi (x)=F\exp \left( -\frac{1}{1-q}\int _{0}^{x}\frac{\ln \left\{ 1-(1-q)t\left[ T(t)-u_{0}(qt)S(t)\right] \right\} }{t}d_{q}t\right) +
\]
\[
+u_{0}(x)\exp \left( -\frac{1}{1-q}\int _{0}^{x}\frac{\ln \left\{ 1-(1-q)t\left[ R(t)+u_{0}(t)S(t)\right] \right\} }{t}d_{q}t\right) \times 
\]
\[
\times \left[ D+F \int _{0}^{x}\frac{S(t)}{1-\left( 1-q\right) t\left[ R(t)+u_{0}(t)S(t)\right] }\times \right. 
\]
\begin{equation}
\label{27a}
\times \left.  \exp \left( \frac{1}{1-q}\int _{0}^{t}\frac{1}{s}\ln \frac{1-\left( 1-q\right) s\left[ R(s)+u_{0}(s)S(s)\right] }{1-\left( 1-q\right) s\left[ T(s)-u_{0}(qs)S(s)\right] }d_{q}s\right) d_{q}t\right]\; ,
\end{equation}
The constants \(  D \) and \(  F \)are related with the initial conditions
by:
\[
D=\psi (0)\: ,
\]
\begin{equation}
\label{28}
F=-\psi (0) \frac{\varphi _{0}(0)}{\psi _{0}(0)}+\varphi (0).
\end{equation}
\end{proposition}

\vspace*{0.2cm}\hspace*{-0.7cm}
{\it Proof:}\\
In order to prove the formulae (\ref{27}- \ref{27a}) and (\ref{28}) we assume
in (\ref{10}) that \(  \Lambda ^{'}(x) \) is uppertriangular and we
apply the $q$--difference Darboux transform (\ref{16a}) with 
\begin{equation}
\label{28a}
\: D(x)=\left( \begin{array}{cc}
1 & 0\\
-u_{0}(x) & 1
\end{array}\right) .
\end{equation}
The formulae (\ref{26}) follows now from (\ref{16a}) and from the property
of \( \: \Lambda ^{'}(x) \) of being uppertriangular. The formulae (\ref{27})
and (\ref{27a}) are obtained from {\bf Proposition 1} and (\ref{28}).

\ \hfill QED

\vspace*{0.2cm}

As a corollary of {\bf Proposition 2} we obtained the auto-B{\"a}cklund transform
for $q$--difference Riccati equation.

\vspace*{0.2cm}

\begin{proposition}
Let \( \: u_{0}(x) \) be some special solution of the equation (\ref{4}).
Then the general solution of (\ref{4}) is given by
\[
u^{t}(x)=\left({\bf\cal B}_{t}^{+}u_{0}\right)(x)=u_{0}(x)+
\]
\begin{equation}
\label{30a}
+\frac{t\exp \left( \frac{1}{1-q} \int _{0}^{x} \frac{1}{y} \ln \frac{1-(1-q)y\left[ R(y)+u_{0}(y)S(y)\right] }{ 1-(1-q)y\left[T(y)-u_{0}(qy)S(y)\right] } d_{q}y\right) }{ 1+t\int _{0}^{x}\frac{S(y)}{1-(1-q)y\left[R(y)+u_{0}(y)S(y)\right] }\exp \left(\frac{1}{1-q}\int _{0}^{y} \frac{1}{s}\ln \frac{1-(1-q)s\left[R(s)+u_{0}(s)S(s)\right]}{1-(1-q)s\left[T(s)-u_{0}(qs)S(s)\right] }d_{q}s\right)d_{q}y },
\end{equation}
where $t\in {I\!\! R}$.
\end{proposition}

\vspace*{0.2cm}\hspace*{-0.7cm}
{\it Proof:}\\
The formulae (\ref{30a}) is obtained by substituting (\ref{27}) and (\ref{27a}) into (\ref{6}) and puting $t=\frac{F}{D}$.

\ \hfill QED

\vspace*{0.2cm}

The $q$--difference Schr{\"o}dinger equation 
\begin{equation}
\label{30b}
\left( -\partial ^{2}_{q}+V(x)\right) \psi (x)=0,
\end{equation}
is a special case of (\ref{4}) and is obtained by putting \( \: R(x)=T(x)=0 \)
and \( \: S(x)=1 \). From the {\bf Proposition 2}  we may draw:

\vspace*{0.2cm}

\begin{corollary}
The solution of $q$--difference Schr{\"o}dinger equation with the potential
\begin{equation}
\label{30}
V(x)=\partial _{q}u_{0}(x)+u_{0}(x)u_{0}(qx),
\end{equation}
is given by
\[
\psi (x)=\exp \left( -\frac{1}{1-q}\int _{0}^{x}\frac{\ln \left( 1-\left( 1-q\right) tu_{0}(t)\right) }{t}d_{q}t\right) \times 
\]
\begin{equation}
\label{31}
\times \left[ D+F\int _{0}^{x}\frac{1}{1-\left( 1-q\right) tu_{0}(t)}\right. \left. \exp \left( \frac{1}{1-q}\int ^{t}_{0}\frac{1}{s}\ln \frac{1-(1-q)su_{0}(s)}{1+(1-q)su_{0}(qs)}d_{q}s\right) d_{q}t\right] \quad .
\end{equation}
\end{corollary}

\vspace*{0.2cm}

In the limit of \( q\rightarrow 1 \) the $q$--difference equations (\ref{5}) and
(\ref{4}) tend to their differential counterparts
\begin{equation}
\label{32}
\frac{d}{dx}\left( \begin{array}{c}
\psi (x)\\
\varphi (x)
\end{array}\right) =\left( \begin{array}{cc}
R(x) & S(x)\\
V(x) & T(x)
\end{array}\right) \left( \begin{array}{c}
\psi (x)\\
\varphi (x)
\end{array}\right) 
\end{equation}
and
\begin{equation}
\label{33}
V(x)=\frac{d}{dx}u(x)+\left( R(x)-T(x)\right) u(x)+S(x)u^{2}(x),
\end{equation}
where \( \: u(x) \) is given by (\ref{6}). The {\bf Propositions 2} and {\bf 3} are
valid in the limit \( \: q\rightarrow 1 \) too. Therefore, we can apply them
to the differential case and reproduce in such a way the formulae for general
solutions of (\ref{32}) and (\ref{33}). They are  given by
\[
\psi (x)=\exp \left( \int _{0}^{x}\left( R(t)+u_{0}(t)S(t)\right)dt \right) \times 
\]
\begin{equation}
\label{34}
\times \left[ D+F\int _{0}^{x}S(t)\exp \left( \int _{0}^{t}\left[ T(s)-R(s)-2u_{0}(s)S(s)\right] ds\right) dt\right] \: ,
\end{equation}
\[
\varphi (x)=F\exp \left( \int _{0}^{x}\left( T(t)-u_{0}(t)S(t)\right) dt\right) +
\]
\[
+u_{0}(x)\exp \left( \int _{0}^{x}\left( R(t)+u_{0}(t)S(t)\right) dt\right) \times 
\]
\begin{equation}
\label{34a}
\times \left[ D+F \int _{0}^{x}S(t)\exp \left( \int _{0}^{t}\left[ T(s)-R(s)-2u_{0}(s)S(s)\right] ds\right) dt \right] 
\end{equation}
and by
\begin{equation}
\label{35}
u^{t}(x)=u_{0}(x)+\frac{t\exp \left(\int _{0}^{x} \left[T(y)-R(y)-2u_{0}(y)S(y)\right]dy\right)}{1+t\int _{0}^{x}S(y)\exp \left(\int _{0}^{y} \left[ T(s)-R(s)-2u_{0}(s)S(s)\right]ds\right)dy}\:.
\end{equation}
Here \( \: \left( \begin{array}{cc}
\psi _{0}(x) & \varphi _{0}(x)
\end{array}\right) ^{\top } \)and \( \: u_{0}(x) \) are some special solutions (\ref{32}) and (\ref{33}).

In order to describe the properties of the family of solutions $u^{t}(x)=\left({\bf\cal B}_{t}^{+}u_{0}\right)(x)$, $t\in {I\!\! R}$, 
given by (\ref{30a}) let us formulate:\\

\vspace*{0.3cm}\hspace*{-0.7cm}
{\bf Proposition 4}\\

\vspace*{-0.4cm}
{\it \begin{itemize}
\item  The transforms (\ref{30a}) form one--parameter group
\begin{equation}
\label{136}
{\bf\cal B}_{t_{1}}^{+}\circ {\bf\cal B}_{t_{2}}^{+}={\bf\cal B}^{+}_{t_{1}+t_{2}},
\end{equation}
which acts transitively on the space of all solutions of the $q$--difference Riccati \mbox{equations (\ref{4}).}\\
\item The solutions $u^{t_{1}}(x)$, $u^{t_{2}}(x)$, $u^{t_{3}}(x)$, $u^{t_{4}}(x)$ satisfy 
the unharmonical superposition principle
\begin{equation}
\label{138}
\frac{\left( u^{t_{4}}(x)-u^{t_{3}}(x)\right) \left( u^{t_{1}}(x)-u^{t_{2}}(x)\right) }{\left( u^{t_{3}}(x)-u^{t_{1}}(x)\right) \left( u^{t_{2}}(x)-u^{t_{4}}(x)\right) }=
\frac{\left(t_{4}-t_{3}\right)\left(t_{1}-t_{2}\right)}{\left(t_{3}-t_{1}\right)\left(t_{2}-t_{4}\right)},
\end{equation}
for $t_{1}$, $t_{2},t_{3},t_{4}\in{I\!\! R}$.
\end{itemize}
}

\vspace*{0.2cm}\hspace*{-0.7cm}
{\it Proof:}\\
$\bullet$ According to {\bf Proposition 3} any solution of (\ref{4}) is given by ${\bf\cal B}_{s}u_{0}$ 
for some $s\in{I\!\! R}$. Since ${\bf \cal B}_{t_{1}}\left({\bf\cal B}_{t}u_{0}\right)$ is 
a solution of (\ref{4}) we have 
\begin{equation}
\label{137}
{\bf\cal B}^{+}_{t_{1}}\circ {\bf\cal B}^{+}_{t_{2}}u_{0}={\bf\cal B}^{+}_{s}u_{0}.
\end{equation}
One can find $s\in{I\!\! R}$ by evaluation of both sides of (\ref{137}) at $x=0$. Thus, using (\ref{28})
we obtain
\begin{equation}
t_{1}+\left(t_{2}+u_{0}(0)\right)=s+u_{0}(0).
\end{equation}
$\bullet$ The equality is obtained from (\ref{30a}) by direct calculation.
\ \hfill QED

\vspace*{0.4cm}

The right hand side of (\ref{138}) is invariant with respect to the real fractional transformation
\begin{equation}
\label{139}
t_{i}^{'}=\frac{at_{i}+b}{ct_{i}+d},\;\;\;
\left(\begin{array}{cc}
a & b \\
c & d 
\end{array}
\right)\in SL(2,I\!\!R)
\end{equation} 
of the parameters $t_{1}$, $t_{2}$, $t_{3}$ and $t_{4}$. Hence, the left hand side of (\ref{138}) is  
$SL(2,I\!\!R)$--invariant too.

\vspace*{0.2cm}

\section{Some integrable cases}

In the {\bf Proposition 3} we constructed the transform which generated the general solution of 
the $q$--Riccati equation (\ref{4}) from a given special one. It was done by the action of one--parameter group
of transformations $\{ {\bf\cal B}^{+}_{t}\}$, $t\in {I\!\!R}$. The formula (\ref{30a}) which defines the group
$\{ {\bf\cal B}^{+}_{t}\} _{t\in {I\!\! R}}$ action does depend on the potentials $R(x)$, $S(x)$ and $T(x)$. 
It does not contain the potential $V(x)$. This allows us to define some method of creation of new 
integrable systems from a system which one knows how to integrate. In order to do this let us intoduce some 
notation. 

By ${\bf\cal I}$ we will denote the map
\begin{equation}
\label{111}
{\bf\cal I}u(x):=-u(x).
\end{equation}
The operator which acts on the function $u(x)$ on the right side of $q$--Riccati equation (\ref{4}) will be 
denoted by ${\bf\cal R}_{+}$, i.e.
\begin{equation}
\label{112}
{\bf\cal R}_{+}u(x):=\partial_{q}u(x)-T(x)u(x)+R(x)u(qx)+S(x)u(x)u(qx).
\end{equation}
By ${\bf\cal R}_{-}$ we will denote the operator
\begin{equation}
\label{113}
{\bf\cal R}_{-}u(x):=-\partial_{q}u(x)+T(x)u(x)-R(x)u(qx)+S(x)u(x)u(qx).
\end{equation}
It is clear that
\begin{equation}
\label{114}
{\bf\cal R}_{+}\circ{\bf\cal I}={\bf\cal R}_{-}\;\;\mbox{\rm and}\;\;
{\bf\cal R}_{-}\circ{\bf\cal I}={\bf\cal R}_{+}.
\end{equation}
Hence, if $u(x)$ is the solution of the equation
\begin{equation}
\label{115}
{\bf\cal R}_{+}u(x)=V(x),
\end{equation}
then ${\bf\cal I}u(x)$ does satisfy
\begin{equation}
\label{116}
{\bf\cal R}_{-}\circ{\bf\cal I}u(x)=V(x)
\end{equation}
and vice versa. The one--parameter groups (\ref{30a}) for the equations 
(\ref{115}) and (\ref{116}) will be denoted by 
$\{ {\bf\cal B}_{t}^{+}\}_{t\in {I\!\! R}}$ and $\{ {\bf\cal B}_{t}^{-}\}_{t\in {I\!\! R}}$
respectively. They are related by
\begin{equation}
\label{117}
{\bf\cal B}_{t}^{-}={\bf\cal I}\circ{\bf\cal B}_{t}^{+}\circ{\bf\cal I}.
\end{equation}
After aplication of the transform
\begin{equation}
\label{118}
{\bf\cal B}^{-}_{t_{1}\ldots t_{n}}:={\bf\cal I}\circ{\bf\cal B}_{t_{1}}^{+}\circ{\bf\cal I}\circ{\bf\cal B}_{t_{2}}^{+}\circ\cdots\circ{\bf\cal I}\circ{\bf\cal B}_{t_{n}}^{+}\circ{\bf\cal I}
\end{equation}
to $u_{0}(x)$ which one assumes to be the solution of (\ref{4}) with the potential $V_{0}(x)$ 
we obtain the solution
\begin{equation}
\label{119}
u(t_{1},\ldots ,t_{n},x):={\bf\cal B}^{-}_{t_{1}\ldots t_{n}}u_{0}(x)
\end{equation}
of (\ref{115}) with the some new potential
\begin{equation}
\label{120}
{\bf\cal R}_{+}u(t_{1},\ldots ,t_{n},x)=V(t_{1},\ldots ,t_{n},x),
\end{equation}
which is $n$--parameter deformation of the initial one. The same function satisfies also
\begin{equation}
\label{121}
{\bf\cal R}_{-}u(t_{1},\ldots ,t_{n},x)=V(t_{2},\ldots ,t_{n},x).
\end{equation}
In such a way we obtain the family of $q$--difference Riccati equations (\ref{120}) 
and (\ref{121}) generated by $u_{0}(x)$ and by the transform (\ref{118}).

It is worth to mention here the group like property of ${\bf\cal B}_{t_{1}\ldots t_{n}}$:
\begin{eqnarray}
\label{122}
{\bf\cal B}^{-}_{0\ldots 0}=id\nonumber ,\\
\left({\bf\cal B}^{-}_{t_{1}\ldots t_{n}}\right)^{-1}={\bf\cal B}_{-t_{n}\ldots -t_{1}},\\
{\bf\cal B}^{-}_{t_{1}\ldots t_{n}}\circ{\bf\cal B}^{-}_{s_{1}\ldots s_{m}}={\bf\cal B}^{-}_{t_{1}\ldots t_{n-1}t_{n}+s_{1}s_{2}\ldots s_{m}} \nonumber .
\end{eqnarray}

In particular of case $n=1$, applying $ \{ {\bf\cal B}^{-}_{t}\}_{t\in {I\!\! R}}$ to (\ref{30}) we obtain 
\begin{equation}
\label{123}
{\bf\cal R}_{+}u(t,x)=V(t,x),
\end{equation}
\begin{equation}
\label{124}
{\bf\cal R}_{-}u(t,x)={\bf\cal R}_{-}u_{0}(x),
\end{equation}
where $u(t,x)={\bf\cal B}^{-}_{t}u_{0}(x)$. Simple calculation gives a solution
\[
u(t,x)=u_{0}(x)-
\]
\begin{equation}
\label{125}
-\frac{t\exp \left( \frac{1}{1-q} \int _{0}^{x} \frac{1}{y} \ln \frac{1+(1-q)y u_{0}(y)}{ 1-(1-q)yu_{0}(qy) } d_{q}y\right) }{ 1+t\int _{0}^{x}\frac{1}{1+(1-q)yu_{0}(y)}\exp \left(\frac{1}{1-q}\int _{0}^{y} \frac{1}{s}\ln \frac{1+(1-q)su_{0}(s)}{1-(1-q)su_{0}(qs)}d_{q}s\right)d_{q}y },
\end{equation}
for the potential 
\[
V(t,x)=V_{0}(x)-
\]
\begin{equation}
\label{126}
-2\partial_{q}\frac{t\exp \left( \frac{1}{1-q} \int _{0}^{x} \frac{1}{y} \ln \frac{1+(1-q)y u_{0}(y)}{ 1-(1-q)yu_{0}(qy) } d_{q}y\right) }{ 1+t\int _{0}^{x}\frac{1}{1+(1-q)yu_{0}(y)}\exp \left(\frac{1}{1-q}\int _{0}^{y} \frac{1}{s}\ln \frac{1+(1-q)su_{0}(s)}{1-(1-q)su_{0}(qs)}d_{q}s\right)d_{q}y }.
\end{equation}
In the limit of $q\rightarrow 1$ the equations (\ref{123}) and (\ref{124}) 
correspond to the proper differential equations and (\ref{125}) tends to the  solution
\begin{equation}
\label{127}
u(t,x)=u_{0}(x)-\frac{\partial}{\partial x}\ln \left( 1+t\int_{0}^{x}\exp\left( 2\int_{0}^{y}u_{0}(z)dz\right) dy\right)
\end{equation}
of (\ref{2}) with potential given by 
\begin{equation}
\label{128}
V(t,x)=V_{0}(x)-2\frac{\partial^{2}}{\partial x^{2}}\ln \left( 1+t\int_{0}^{x}\exp\left( 2\int_{0}^{y}u_{0}(z)dz\right) dy\right).
\end{equation}

Combining (\ref{123}) and (\ref{124}) we find the formula
\[
u(t,x)=\frac{1}{2}\left\{ \left( 1-q \right) x \frac{1}{2}\left( V(t,x)-{\bf\cal R}_{-}u_{0}(x)\right) \stackrel{+}{-}\right. 
\]
\begin{equation}
\label{127AA}
\left. \stackrel{+}{-}\sqrt{\left[ \left(1-q\right) x\frac{1}{2}\left(V(t,x)-{\bf\cal R}_{-}u_{0}(x)\right)\right]^{2}+2\left( V(t,x)+{\bf\cal R}_{-}u_{0}(x)\right)}\right\} ,
\end{equation}
which expresses the solution $u(t,x)$ by the $t$--deformed potential $V(t,x)$ 
and by the initial solution $u_{0}(x)$.

\vspace*{0.4cm}
{\bf Example}\\
The function
\begin{equation}
u_{0}(x)=ax^{\alpha},
\end{equation}
where $\alpha>-1$,  $a,\alpha\in{I\!\! R}$, is the solution of (\ref{30}) 
with the potential
\begin{equation}
\label{129}
V_{0}(x)=a\frac{1-q^{\alpha}}{1-q}x^{\alpha-1}+a^{2}q^{\alpha}x^{2\alpha}.
\end{equation}
Let
\begin{equation}
\label{d}
\exp_{R}(x)=\sum_{n=0}^{\infty}\frac{1}{R(q)\cdots R(q^{n})}x^{n},
\end{equation}
be generalized exponential function (see \cite{3}) with
\begin{equation}
\label{131}
R(x)=\frac{1-x^{\alpha+1}}{\left(1-q\right)x^{\alpha+1}}.
\end{equation}
Using the formulae (\ref{125}) and (\ref{126}) we find that
\begin{equation}
\label{130}
u(t,x)=ax^{\alpha}-\frac{t\frac{\exp_{R}\left( ax^{\alpha +1} \right) }
{\exp_{R}\left( -aq^{\alpha}x^{\alpha +1}\right)}}
{1+t\int_{0}^{x}\frac{\exp_{R}\left( aq^{\alpha+1}y^{\alpha+1}\right)}
{\exp_{R}\left(-aq^{\alpha}y^{\alpha+1}\right)}d_{q}y},
\end{equation}
is a solution of (\ref{30}) with potential $V(t,x)$ given by
\[
V(t,x)=a\frac{1-q^{\alpha }}{1-q}x^{\alpha-1}+a^{2}q^{\alpha}x^{2\alpha}-
\]
\[
-2t\frac{
\frac{\exp_{R}\left( aq^{\alpha +1}x^{\alpha +1} \right) }
{\exp_{R}\left( -aq^{\alpha}x^{\alpha +1}\right)}}
{\left( 1+t\int_{0}^{x} 
\frac{\exp_{R}\left( aq^{\alpha+1}y^{\alpha+1}\right)}
{\exp_{R}\left( -aq^{\alpha}y^{\alpha+1}\right)}d_{q}y\right)
\left( 1+qt\int_{0}^{x} \frac{\exp_{R}\left( aq^{2(\alpha+1)}y^{\alpha+1}\right)}
{\exp_{R}\left( -aq^{2\alpha+1}y^{\alpha+1}\right)}d_{q}y\right)}\times
\]
\begin{equation}
\label{132}
\times
\left\{ \left(1+q^{\alpha}\right) aq^{\alpha+1}x^{\alpha}
\left( 1+t\int_{0}^{x}\frac{\exp_{R}\left( aq^{\alpha+1}y^{\alpha+1}\right)}{\exp_{R}\left(-aq^{\alpha}y^{\alpha+1}\right)}d_{q}y\right)
-t\frac{\exp_{R}\left( aq^{\alpha +1}x^{\alpha +1} \right) }
{\exp_{R}\left( -aq^{2\alpha +1}x^{\alpha +1}\right)}\right\}.
\end{equation}

If, for example, $\alpha =0$, $t=a$ and $q\rightarrow 1$ then  the potential (\ref{132}) reduces to Rosen--Morse
potential
\begin{equation}
V(a,x)=a^{2}\frac{\exp\left(-2ax\right)-6 +\exp\left( 2ax\right)}{\left(\exp\left( -ax\right) +\exp\left( ax\right)\right)^{2}}.
\end{equation}

\vspace*{0.5cm}

\begin{center}

{\bf Acknowledgements}

\end{center}

We thank dr Z Hasiewicz for helpful comments.

\end{document}